\documentclass[aps,prb,twocolumn,showpacs]{revtex4}
\usepackage{graphicx}
\usepackage{epsf}
\begin{document}

\newcommand{\figureheight}{8.2 cm}
\newcommand{\putfig}[2]{\begin{figure}[h]
        \special{isoscale #1.bmp, \the\hsize \figureheight}
        \vspace{\figureheight}
        \caption{#2}
        \label{fig:#1}
        \end{figure}}

\newcommand{\eqn}[1]{(\ref{#1})}
\newcommand{\be}{\begin{equation}}
\newcommand{\ee}{\end{equation}}
\newcommand{\bea}{\begin{eqnarray}}
\newcommand{\eea}{\end{eqnarray}}
\newcommand{\bean}{\begin{eqnarray*}}
\newcommand{\eean}{\end{eqnarray*}}
\newcommand{\nn}{\nonumber}




\title{Modulation of Luttinger liquid exponents in multi-walled
carbon nanotubes}
\author{S. Bellucci$^a$, J. Gonz\'{a}lez$^b$, P. Onorato$^{a,c}$
and E. Perfetto$^{a,b}$ \\}
\address{
        $^a$INFN, Laboratori Nazionali di Frascati,
        P.O. Box 13, 00044 Frascati, Italy. \\
        $^b$Instituto de Estructura de la Materia,
        Consejo Superior de Investigaciones Científicas, Serrano 123,
28006 Madrid, Spain\\
        $^c$Dipartimento di Scienze Fisiche,
        Universit\`{a} di Roma Tre, Via della Vasca Navale 84,
00146 Roma, Italy} 
\date{\today}
\begin{abstract}
We develop in this paper a theoretical framework that applies to
the intermediate regime between the Coulomb blockade and the
Luttinger liquid behavior in multi-walled carbon nanotubes.
Our main goal is to confront the experimental observations of
transport properties, under conditions in which the thermal
energy is comparable to the spacing between the single-particle
levels. For this purpose we have devised a many-body approach to
the one-dimensional electron system, incorporating the effects
of a discrete spectrum. We show that, in the crossover regime,
the tunneling conductance follows a power-law behavior as a
function of the temperature, with an exponent that oscillates
with the gate voltage as observed in the experiments. Also in
agreement with the experimental observations, a distinctive
feature of our approach is the existence of an inflection
point in the log-log plots of the conductance vs temperature,
at gate voltages corresponding to peaks in the oscillation
of the exponent. Moreover, we evaluate the effects of a
transverse magnetic field on the transport properties of the
multi-walled nanotubes. For fields of the order of 4 T,
we find changes in the band structure that may be already
significant for the outer shells, leading to an appreciable
variation in the power-law behavior of the conductance. We
then foresee the appearance of sensible modulations in the
exponent of the conductance for higher magnetic fields, as
the different subbands are shifted towards the development
of flat Landau levels.

\end{abstract}

\pacs{73.63.Fg, 73.22.-f, 73.23.-b}

\maketitle

\section{Introduction}
Recent progresses in nanotechnology has revealed many novel
transport phenomena in mesoscopic low-dimensional structures.
Carbon nanotubes (CNs) were discovered by Iijima in 1991\cite{1},
as a by-product of carbon fullerene production. Thus, since then,
a new field of research in mesoscopic physics has opened\cite{cbmac}.
CNs are basically rolled up sheets of graphite (hexagonal
networks of carbon atoms) forming tubes that are only nanometers
in diameter. The electronic properties of CNs depend on
their diameter and chiral angle (helicity) parameterized by a
roll-up (wrapping) vector $(n, m)$\cite{3n}. Hence it follows that
some nanotubes are metallic with high electrical conductivity, while
others are semiconducting with relatively low band gaps.
They may also display different electronic properties depending on
whether they are single-walled carbon nanotubes (SWNTs) or
multi-walled carbon nanotubes (MWNTs). MWNTs are typically made of
several (typically 10) concentrically arranged graphene sheets with
a radius of about $5 \; {\rm nm}$ and lengths in the range of
$1-100 \; {\rm \mu m}$.

Several electrical transport experiments have been done by
using CNs. In these experiments the effects of the Coulomb interaction
are quite relevant, and the way they manifest depend mainly
on the size of the system, the temperature, and the quality of the
contacts used in experiments.
Thus, MWNTs show for instance different transport regimes
from ballistic transport (conductance quantization\cite{10,11},
Luttinger liquid-like\cite{12,13}) to diffusive transport\cite{ba1}
and also strong localization\cite{ka1,241}.

When the contacts are not highly transparent, the measurements of the
conductance and the differential conductivity reflect the strong
Coulomb repulsion in CNs\cite{e1,e2}. At low temperature,
$T\lesssim 1$ K, the zero-bias conductance is strongly influenced
by quantum mechanical effects and  shows regular oscillations
corresponding to the so-called \emph{single electron tunneling} (SET),
because it arises from the electrons going one at a time through
the device\cite{e3,e4}. The voltage between two peaks is
determined by the energy for adding an  electron to the device
between the potential barriers. This addition  energy depends on
the level spacing, due to the momentum quantization,
and on the electron-electron interaction\cite{noicb}. This is
the Coulomb blockade (CB) regime where the tunneling transparency
of a barrier vanishes owing to the electron-electron interaction.
The differential conductivity of individual MWNTs has been measured in
Ref. \onlinecite{bui}, where CB and energy level quantization have been
observed.

When the temperature increases, so that the thermal
energy is much larger than the level spacing, the transport
measurements reflect instead the many-body properties of the
system. In fact, the electron-electron interaction in a
one-dimensional (1D) system
is expected to lead to the formation of a Luttinger liquid (LL)
with properties very different from those of the non-interacting
Fermi gas\cite{TL}. The LL state is characteristic of
1D clean (ballistic) conductors, which have unusual
electronic properties that cannot be explained by Fermi liquid
theory.

Landau quasiparticles are unstable in 1D systems of
interacting electrons, in which the low-energy excitations
take the form of plasmons (collective
electron-hole pair modes) that are the stable excitations: this is
known as the  breakdown of Fermi liquid in 1D\cite{4a,sol,hald}.
The LL state has two main features:

i) the power-law dependence of physical quantities, such as the
tunneling
density of states (TDOS), as a function of energy or temperature;

ii) the spin-charge separation: an additional electron in the LL
decays into decoupled spin and charge wave packets, with different
velocities for charge and spin.

Characteristic experimental signatures support the assumed LL
behaviour of CNs\cite{48,48b,48c,48d}. In fact the power-law
dependence of physical observables follows from the behavior of
the TDOS as a function of the energy or
the temperature: the tunneling conductance $G$ reflects the power
law dependence of the TDOS in  a small-bias experiment\cite{kf}
\bea
 G=dI/dV\propto
T^{\alpha }
\eea
for $e V_b \ll k_B T$, $V_b$ being bias voltage.
Such kind of signatures of LL behavior
have been observed in single-walled as well as in multi-walled
nanotubes. In MWNTs the observed values of $\alpha $ lie
between $0.04$ and $0.3$, and results in agreement with the
experimental values have been obtained in previous theoretical
works\cite{noi1,noi2,npb,noi4}.

In this paper we introduce a theoretical description of  the
crossover as the temperature decreases from the LL
regime to the CB regime. We do this in order to confront
the experimental data reported in a recent letter by Kanda {\em et
al.}\cite{kprl}, in which the intermediate regime has been explored
measuring the zero-bias conductance at temperatures where the
thermal energy becomes comparable to the level spacing in the
discrete single-particle spectrum.
In Ref. \onlinecite{kprl} the authors have reported a systematic
study of the gate voltage dependence of the LL-like behaviour in
MWNTs, showing the dependence of the exponent $\alpha $ on
gate voltage, with values of $\alpha$ ranging from $0.05$ to $0.35$.
The main results of Ref. \onlinecite{kprl} are as follows:

i) the gate-voltage ($V_g$) dependence of the exponent $\alpha$
below $30$ K exhibits periodic oscillations; the characteristic
$V_g$ scale for $\alpha$ variation, $\Delta V_g$, is around $1 V$;

ii) changes in the exponent $\alpha$ are observed in the plots of
the conductance
at an inflection temperature $T^* \sim 30$ K, for values of $V_g$
corresponding to peaks of $\alpha$.

iii) the exponent $\alpha$ depends significantly on the transverse
magnetic field acting on the CN. $\alpha$ is reduced from a
value of $0.34$, at a peak in the $\alpha$ oscillations,
to a value $0.11$, for a magnetic field $B = 4$ T. In most cases,
the exponent $\alpha $ becomes smaller for higher magnetic fields.

In their letter, Kanda {\em et al.} do not find plausible
the explanation of the mentioned features starting from a
LL description of the MWNTs. In this respect, Egger has considered
in Ref. \onlinecite{eg99} a model for MWNTs composed of a number
$N_{SH}$ of ballistic metallic shells. The author has discussed
there a low-energy theory for the MWNTs including long-ranged
Coulomb interactions and internal screening effects.
The theory may be also extended to include the effect of a
variable number of conducting modes modulated by the doping
level. However, Kanda {\em et al.} rule out the possibility that
the change in the number of subbands at the Fermi level may be
at the origin of the features observed in their experiment, as
long as the subband spacing is too large to be consistent with
the period of the oscillations.

Then, Kanda {\em et al.} turn to a different kind of theory
that considers the MWNT as a diffusive conductor. Egger and
Gogolin\cite{eg01} have calculated the TDOS of doped MWNTs
including disorder (with mean free path $l$ smaller than the
radius $R$) and
electron-electron interactions. MWNTs may display an effective
and nonconventional CB arising from tunneling into a strongly
interacting disordered metal, leading to LL-like zero-bias
anomalies: the exponent becomes
$\alpha=(R/2\pi \hbar D \nu_0)\log(1+\nu_0U)$,
where $D$ is the diffusion constant,
$\nu_0= N_s/2\pi \hbar v_F$ is the noninteracting density of
states depending on the number of subbands $N_s$ and the Fermi
velocity $v_F$, and $U_0$ is an effectively short-ranged 1D
interaction. Substituting these parameters by pertinent values
yields an estimate $\alpha\simeq R/N_sl$, that is near the
experimentally observed values.

Kanda and coworkers conclude that this second theory better
explains the experimental results, by assuming that the  mean-free
path $l$  may fluctuate with the gate voltage. This is based on
the theoretical work by Choi {\em et al.}\cite{ch}, that
have studied the effects of single defects on the local density
of states via resonant backscattering. However, the argument of
Kanda {\em et al.} has not addressed the question of how a
random distribution of defects may produce the oscillations
observed in the experiment. We believe otherwise that the
depedence of the $\alpha $ exponent on the gate voltage is in
correspondence with a definite periodic structure of the
single-particle density of states.

In the next section, we put forward the idea that there exists
a clear relation between the periodic oscillations in the
$\alpha $ exponent and the quantization of the energy levels
in the MWNTs\cite{noiprl}.
Building on this idea, in section III we introduce
an energy distribution function which takes into account the effects
of the quantization at low temperature and can be compared with
the Fermi-Dirac distribution in the limit of high temperature.
Next, in section IV we develop a theoretical approach based on
many-body theory, in order to deal with the low-energy
effects of the the long-range Coulomb interaction in the 1D electron
system. We introduce in that scheme the discussed energy distribution
function, for the sake of evaluating the effects of the discrete
spectrum on the exponent $\alpha$. The results of this approach are
reported in section V, where we carry out the comparison between our
theoretical predictions and the experiments. The effects of a
transverse magnetic field are considered in section VI. In section
VII we highlight how the distinctive features of our approach match
naturally with the experimental observations, and we close with
some concluding remarks and perspectives in section VIII.

\section{ANALYSIS OF THE EXPERIMENTAL OSCILLATIONS}

Here we discuss the values of relevant parameters, in order to
compare later the experimental results with the theoretical
predictions.

The transverse quantization energy of the nanotube, which
corresponds to the distance between two nearest subbands, is
given by $\Delta E_\perp \approx \hbar v_F/R$, where the Fermi
velocity is $v_F = 8 \times 10^5 \; {\rm m/s}$ and we take a
nanotube radius $R \approx 5 \; {\rm nm}$. On the other hand, the
longitudinal quantization energy reads $\Delta \varepsilon_L=
hv_F/L$, where in the experiments $L\approx 4\div6 \mu m$
(here we are considering the entire length of the MWNT,
$L\approx 5.7 (4.7) \; {\rm \mu m}$, rather than the length of its
portion between the electrodes, $d\approx 0.7 (1.0) \; {\rm \mu m}$).
The ratio  turns out to be $\frac{\Delta E_\perp}{\Delta
\varepsilon_L}\approx \frac{L}{2 \pi R}\approx 10 \div  100 $.

In comparing the theoretical estimates with the experiment,
there appears a conversion factor given by the coupling
constant $\gamma$ between the gate electrode and the MWNT. This
factor is approximately $\gamma=C/C_0$, where $C$ is the
capacitance between the metal electrodes and the MWNT while $ C_0$
is the capacitance between the gate electrode and the MWNT. In the
experiments by
Kanda {\em et al.}\cite{kprl,kapl} $C \approx 1 fF$ and $C_0
\approx 1 aF$ so that we obtain $\gamma\approx 1\times 10^3$.

Using these parameters Kanda and coworkers confute the above
mentioned model based on the LL theory. In fact, the transverse
quantization energy is about $0.1 \; {\rm eV}$, so that the typical
value $\Delta V_g=\gamma \Delta E_\perp$ for the change in the
number of subbands $N_s$ at the Fermi level is about $10^2 V$. This
value is much larger than the period for the gate voltage in the
observed oscillations, which corresponds to some Volts. This theory
would also predict a stepwise change in $\alpha$
as a function of $V_g$, which is not in agreement with the
oscillatory behavior measured.

In the experiment, the measured value of the period
$\Delta V_g$ is around 3 V at a temperature $T$ below $30$ K.
As shown by the cited authors in a previous paper\cite{kapl}, where
they reported the observation of the CB effect in
MWNT, a similar value of $\Delta V_g$ corresponds to a period (a
sequence of 4 peaks or diamonds) in the zero bias conductance at
very low temperature $T\sim 30 \; {\rm mK}$.

Following the usual formalism\cite{noicb} we can calculate the
period as $\Delta V_g=\gamma(v_F h/L+4U)$\cite{bui}.
If we neglect the electron-electron interaction $U$, we can
assume $\Delta V_g
\sim \gamma v_F h/L$, in order to  compare the
number of oscillations in a definite range of the gate voltage
($-10 V$ to $10 V$ in the experiment) observed  for two different
nanotubes (i.e. samples A and B in Fig. 1, respectively).
Thus, we expect that the
number of oscillations depends on the length as
$$
\frac{n_A}{n_B}=\frac{L_A}{L_B}=\frac{5.7 \mu m}{4.7\mu m}\approx
1.2.
$$
\begin{figure}
\includegraphics*[width=1.0\linewidth]{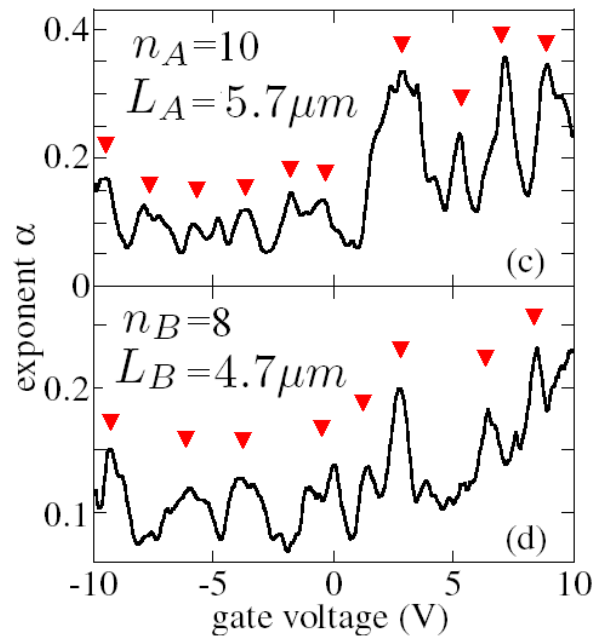}
 \caption{A figure summarizing
the gate-voltage dependence of $\alpha$ in sample A ($L_A=5.7\mu
m$ top) of Ref. \onlinecite{kprl}, calculated from the data below $30^o
K$. A similar $\alpha-V_g$ plot is obtained for sample B ($L_A=4.7\mu m$
bottom).}
\end{figure}

From the experimental data (see Fig. 1) we observe respectively
$10$ and $8$ oscillations, corresponding to an energy spacing
$\Delta \varepsilon \approx 2\div 2.5 \; {\rm meV}$, in perfect agreement
with our estimate. This agreement suggests a theory based on a
microscopic description of the single-particle spectrum,
in order to explain the oscillations. Here
the quantized energy levels play a central role. This theory has
to be valid in the intermediate regime between very low
temperature, where the peaks of the CB are measured by increasing
$V_g$, and room temperature, where the LL behaviour prevails.
To this aim, we will rely on the 1D many-body approach, incorporating
at the same time the discrete character of the spectrum which arises from
the finite geometry of the system.

\section{SINGLE-PARTICLE PROPERTIES}

In the introduction of microscopic features in the many-body formalism,
a central role is played by the energy distribution function (EDF) $n
(\varepsilon ; \varepsilon_F)$.
In the many-body approach to electron liquids, the function
$n(\varepsilon,\varepsilon_F)$ corresponds to the Fermi-Dirac
distribution, which is usually replaced by the Heaveside step function
$\vartheta(\varepsilon-\varepsilon_F)$ in order to simplify the
calculations.

At very low temperatures, as the ones that are needed for the CB
regime, the EDF has to reflect the discreteness of the energy
levels. Thus, some pattern of peaks, corresponding to the quantized
energy levels, has to be present. Hence, at low temperatures, where the
microscopic level quantization is manifest, peaks arise with a shape
smoothed by the growth of the temperature.
The spacing $\Delta \varepsilon$ of the peaks depends on the
nanotube length $L$
$$
 \Delta
\varepsilon =\frac{v_F h}{L},
$$
This value yields a temperature $ T_c \approx
\Delta \varepsilon /k_B $, above which we expect that the
oscillations will be smoothed.

The thermal energy $k_B T_c$ governs then the crossover between two
different regimes. From our analysis, it is clear that two scales
for the behavior as a function of temperature characterize the system,
corresponding to the longitudinal and the transverse quantization
energies, although the second one turns out to be about $10 \div 100$
times larger than $k_B T_c$.

We can model the $n$-th peak of the EDF by a normalized function
$f(\varepsilon, n)$, with a dependence on temperature consistent
with the line shape for a thermally broadened resonance\cite{been}
\begin{equation}
f (\varepsilon, n) = \frac{1}{4k T \cosh^2
  \left( \frac{\varepsilon - n \Delta \varepsilon }
        {2 k T}     \right)},
\end{equation}
where $n \Delta \varepsilon$ stands for the position of the $n$-th
energy level. The energy distribution function is cutoff by the
Fermi energy $\varepsilon_F$, as it is given by
\begin{equation}
n (\varepsilon ; \varepsilon_F ) = \sum_{n=0}^{\infty}
  f (\varepsilon, n ) \, \theta (\varepsilon_{F}-\varepsilon).
\label{edf}
\end{equation}
Alternatively we could introduce a smooth cutoff at the Fermi
energy, taking into account thermal effects
\begin{equation}
n (\varepsilon ; \varepsilon_F ) = \sum_{n=0}^{\infty}
  f (\varepsilon, n ) \, \frac{1}{1+e^{
   ( \varepsilon -\varepsilon_F) /k T }}.
\end{equation}
Anyway the sharp cutoff given by $\theta
(\varepsilon_{F}-\varepsilon)$ is more convenient, in order to
proceed with analytical computations. On the other hand, we have
checked that the presence of the smooth cutoff does not change
appreciably the whole scenario.

At high temperatures, we can neglect the level spacing and
calculate the EDF by integrating
\begin{eqnarray} n
(\varepsilon ; \varepsilon_F )
 & \approx &
    \int_{-\infty}^{\varepsilon_F/\Delta \varepsilon - 1/2}
     dy  f(\varepsilon, y )                 \nonumber \\
  & = & \frac{1}{1+e^{
   ( \varepsilon -\varepsilon_F +\Delta \varepsilon/2 )/k T }}.
\label{cont}
\end{eqnarray}
The latter formula gives the Fermi-Dirac distribution, which is
usually  approximated with the step function. However, the energy
distribution function has in general a more structured shape, that
depends on the ratio between the level spacing $\Delta \varepsilon
$ and the thermal energy $k T$ as illustrated in Fig. 1.

\begin{figure}
\includegraphics*[width=1.0\linewidth]{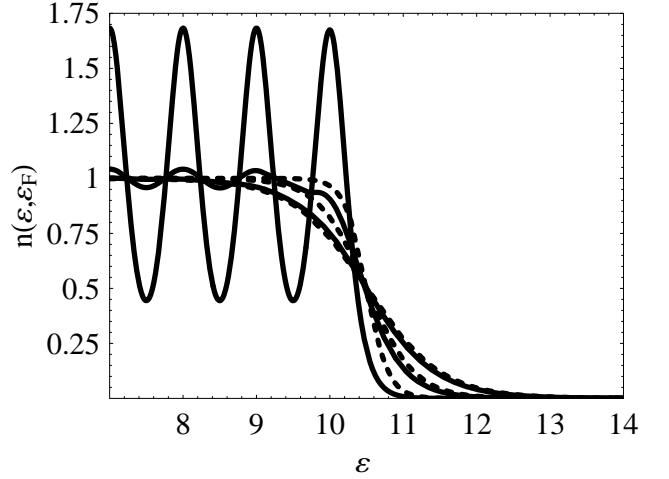}
 \caption{Plot of the energy distribution function $n (\varepsilon
; \varepsilon_F )$ for different values of the temperature
corresponding to $k T/\Delta \varepsilon = 0.25, 0.6, 1.0 $ . The
solid lines have been obtained from Eq. (\ref{edf}) and the dashed
lines by using the Fermi-Dirac distribution (\ref{cont}).}
\end{figure}

The structure of peaks in the energy distribution function affects
several objects in the many-body theory, like the one-loop
polarization $\Pi^{(0)} (q, \omega_q)$. This object counts the
number of particle-hole excitations that can be built with
momentum $q$ across the Fermi level. Thus, in the case of a model
with linear dispersion $\varepsilon (k) \approx v_F k$, it becomes
proportional at zero temperature to
\begin{eqnarray}
\int_{-\infty}^\infty \frac{dk}{2\pi} \: \frac{\theta(q+k-k_F) -\theta(k-k_F)}{v_{F}q} \nonumber \\
= \frac{1}{v_{F}q} \int_{0}^{q} \frac{dk}{2\pi}  = \frac{1}{2\pi
v_{F}} . \label{norm}
\end{eqnarray}
In the approach based on  the distribution function given by
(\ref{edf}), the peaks give rise to a factor in the integration
that reflects the discrete structure. Therefore Eq.(\ref{norm})
becomes
\begin{eqnarray}
 \int_{-\infty}^\infty \frac{dk}{2\pi} \: \Delta \varepsilon  \frac{n(v_{F}(q+k); \varepsilon_{F}) -n(v_{F}k;\varepsilon_{F})}{v_{F}q} \nonumber \\
= \frac{1}{v_{F}q} \int_{0}^{q} \frac{dk}{2\pi}  \Delta
\varepsilon  \sum_{n=0}^{\infty}
  f (v_{F}k, n ) \approx \frac{1}{2\pi v_{F}} N_{T}(\varepsilon_{F}) \, , \label{fact}
\end{eqnarray}
where at first order the function $N_{T}(\varepsilon_{F})$ can be
taken independent on $q$. Given the sharp cutoff, the $N_{T}$
function gets a simple form:
\begin{eqnarray}
&&\int_{0}^{q} dk \Delta \varepsilon  \sum_{n=0}^{\infty}
  f (v_{F}k, n ) = \nonumber \\
&& \frac{\Delta \varepsilon }{2v_{F}} \sum_{n=0}^{\infty}
[\tanh(\frac{\varepsilon_{F}+v_{F}q -n \Delta \varepsilon}{2kT} )
-\tanh( \frac{ \varepsilon_{F}-n \Delta \varepsilon }{2kT})] \nonumber \\
&&\approx q \frac{\Delta \varepsilon }{4 k T} \sum_{n=0}^{\infty}
[1-\tanh^{2}(\frac{\varepsilon_{F}-n \Delta \varepsilon}{2kT})] =
q  N_{T}(\varepsilon_{F}) \, .
\end{eqnarray}
Therefore it holds
\begin{equation}
N_{T}(\varepsilon_{F}) = \frac{\Delta \varepsilon }{4 k T}
\sum_{n=0}^{\infty} [1-\tanh^{2}(\frac{\varepsilon_{F}-n \Delta
\varepsilon}{2kT})]   \, .
\end{equation}

\begin{figure}
\includegraphics*[width=1.0\linewidth]{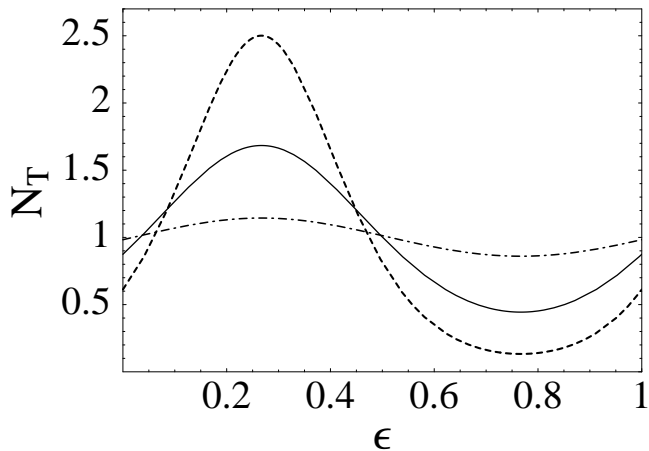}
 \caption{Plot of $N_{T}(\varepsilon_{F})$ for three different values of $ kT / \Delta \varepsilon $:
$ kT / \Delta \varepsilon $ =0.1 (dashed), =0.15 (solid), =0.3
(dotted-dashed). $\varepsilon$ is in unit of $\Delta \varepsilon$
}
\end{figure}
This function is represented in Fig. 3. As expected,
$N_{T}(\varepsilon)$ is periodic with period $\Delta \varepsilon$.

We introduced the function $N_T$ depending on the Fermi energy
$\varepsilon_F$. In this respect, the position of $\varepsilon_F$,
tunable by the gate voltage $V_g$, fixes the portion of the last
peak that we have to integrate. It is clear that, from an
experimental point of view, $\varepsilon_F$ is in correspondence
with the value of the gate voltage $V_g$, so that it can be
shifted in a continuous way, allowing for the exploration of the
region between two peaks in Fig. 2.

\section{MANY-BODY APPROACH}

In the CNs, the relevant interaction is given by the large
Coulomb potential at small momentum-transfer\cite{eg,kane}.
It is known that the Coulomb interaction remains long-ranged in a
1D electron system\cite{13n,noi1}, and this property is also shared
by the graphene sheet, which has a vanishing 2D density of states
at the Fermi level. To obtain a sensible expression of the
1D interaction, we may start from the representation of the
$1/|{\bf r}|$  Coulomb potential in three spatial dimensions as
the Fourier transform of the propagator $1/{\bf k}^2$
\begin{equation}\label{ur}
\frac{1}{|{\bf r}|} =  \int \frac{d^3 k}{(2\pi )^3}
      e^{i {\bf k \cdot r}   }
       \;   \frac{1}{ {\bf k}^2 }
\end{equation}
If the interaction is projected onto one spatial dimension, by
integrating for instance the modes in the transverse
dimensions, then the Fourier transform has the usual logarithmic
dependence on the longitudinal momentum\cite{noi1}
\begin{equation}\label{ux}
\frac{1}{|x|} \approx   - \int \frac{d k}{2\pi } e^{ikx}
       \;   \frac{1}{2\pi } \log (k)
\end{equation}
In our case, the scale of the longitudinal momentum $k$ in the
logarithm is dictated by the existence of a short-distance cutoff
$k_c$, from the small transverse size of the electron system.
Thus, we end up with a representation of the Fourier transform
$\tilde{V} (k)$ of the 1D Coulomb potential
\begin{equation}
\tilde{V} (k) \approx \frac{1}{2\pi } \log (k_c / k)
\end{equation}
where $k_c \sim 1/R$.

Our interest focuses in describing the interaction among a
number $N_s$ of subbands crossing the Fermi level in the
conducting shell of a multi-walled nanotube. Thus, we write
the hamiltonian for the electron fields $\Psi_{\pm a \sigma}$
encoding the modes of the different branches with linear
dispersion about the Fermi points $p_{Fa}, -p_{Fa}$:
\begin{widetext}
\begin{eqnarray}
H & = &  \int_{-k_c}^{k_c } d p \;  \sum_{a = 1}^{N_s}
   \psi_{\pm a \sigma}^{+} (p) \; (\pm v_F p)  \;
            \psi_{\pm a\sigma} (p)
   { \;  + \;    e^2 \int_{-k_c}^{k_c } d p   \;
   \rho (p)   \;  \tilde{V} (p)  \;
          \rho (-p)  \;\;\;\;\;\; }
\label{ham}
\end{eqnarray}
\end{widetext}
In the above expression, $\rho (p)$ stands for the charge
density made from the sum of all the electron densities for the
fields $\psi_{\pm a\sigma} (p)$.

In the hamiltonian (\ref{ham}), we have neglected interactions
that lead to a change of chirality ($\pm $) of the fields, as
they imply a large momentum-transfer of the order of $2k_F$ and
are therefore subdominant with respect to the large Coulomb
interaction at small momentum-transfer. Those so-called
backscattering interactions are known to remain negligible
down to extremely low energies. On the other hand, the model
build from interactions between currents with well-defined
chirality has some interesting properties. One of them is that,
when performing the perturbative calculation of the polarization
function $\Pi (q, \omega_q)$, the self-energy corrections cancel
order by order against the vertex corrections. Thus, the exact
result is just given by the bare polarization funcion. In our case,
upon the introduction of the different subbands and the mentioned
effects in the single-particle density of states, we have
\begin{equation}
\Pi (q, \omega_q) =  \frac{2}{\pi }
  N_s N_T(\varepsilon_F )
 \frac{v_F q^2}{\omega_q^2 - v_F^2 q^2 }
\label{pol}
\end{equation}

\begin{figure}
\includegraphics*[width=1.0\linewidth]{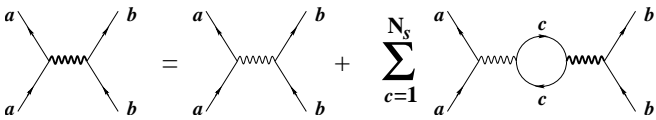}
\caption{Self-consistent diagrammatic equation for the
dressed Coulomb interaction.}
\label{drs}
\end{figure}

Moreover, there is further simplification of the diagrammatics
from the fact that all the fermion loops with more than two
interaction lines vanish identically in 1D. Thus, the dressed
Coulomb interaction has to satisfy the self-consistent
equation shown in Fig. \ref{drs}, that amounts to the RPA sum of
particle-hole contributions.

\begin{figure}[h]
\includegraphics*[width=1.0\linewidth]{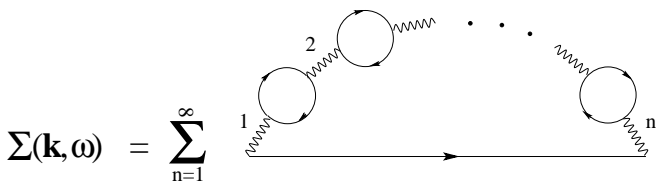}
\caption{Set of diagrams for the computation of the
electron self-energy, where each particle-hole bubble represents
the full polarization operator.}
\label{rain}
\end{figure}

We are mainly interested in obtaining the quasiparticle properties
in the model governed by (\ref{ham}), which requires the
computation of the electron self-energy $\Sigma (q, \omega_q)$.
As long as the RPA provides the exact expression of the dressed
Coulomb interaction, a sensible representation of the self-energy
is given by the partial sum of the perturbative expansion shown
in Fig. \ref{rain}. Correspondingly, we may express the electron
self-energy in the form
\begin{eqnarray}
\Sigma (k, \omega_k) & = & i
      \int_{-k_c}^{k_c}
        \frac{dq}{2\pi } \int_{-\infty}^\infty
        \frac{d \omega_q }{2\pi }
      G^{(0)} (k-q, \omega_k - \omega_q)     \nonumber  \\
   &   &  \times \frac{\tilde{V} (q) }
                  {1 - \tilde{V} (q) \Pi (q, \omega_q) },
\label{selfe}
\end{eqnarray}
where $G^{(0)} (k, \omega_k)$ stands for the free electron
propagator. For the evaluation of the scaling properties of the
quasiparticle parameters, it is enough to obtain the logarithmic
dependence of the function $\Sigma (q, \omega_q)$ on the
high-energy cutoff $E_c \sim v_F k_c$. Thus, the representation in
Fig. \ref{rain} misses corrections to the fermion line $G^{(0)}$
as well as corrections to the three-point vertex at the end of the
line, but these do not introduce new divergences in the cutoff
$E_c$ and can be safely disregarded in the determination of the
electron field scale.

We therefore concentrate on the computation of the contributions
that diverge with the cutoff $E_c$ in (\ref{selfe}). Order by
order in perturbation theory, we find that the same kind of
logarithmic dependence appears in the linear terms proportional to
$k$ and $\omega_k$. We can recast the whole perturbative series
into the expression
\begin{eqnarray}
\Sigma (k, \omega_k)  & \approx &
 \frac{1}{2 \pi^{3/2}}  \frac{\tilde{V} (q_0) e^2}{v_F}
 \sum_{n=0}^{\infty} (-1)^n g^n
    \frac{n \Gamma (n + 1/2)}{(n+1)!}       \nonumber  \\
  &   &  \times     (\omega_k \pm  v_F k) \log (E_c)        \\
  & = &  - \frac{1}{4 N_s N_T }
 \left( \sqrt{1+g} + \frac{1}{\sqrt{1+g}} - 2 \right) \nonumber \\
  &   &  \times     (\omega_k \pm  v_F k) \log (E_c)
\label{sum}
\end{eqnarray}
where the effective coupling is given by
$g = 2 N_s N_T \tilde{V} (q_0)  e^2/\pi v_F$, in terms of a suitable
average momentum $q_0$\cite{foot}.

The frequency and momentum-dependence in Eq. (\ref{sum}) shows
that such self-energy corrections represent a renormalization
of the electron field operator. Then it is convenient to define
a renormalized electron field $\Psi_R $ related in each case to
the bare electron field $\Psi $ through a scale factor
$Z^{1/2} (E_c)$:
\begin{equation}
\Psi_R = Z^{1/2} (E_c) \Psi
\end{equation}
The renormalization factor $Z (E_c)$ must be fixed so that it
makes finite the renormalized electron Green function
$G(k, \omega_k)$. This is given by
\begin{eqnarray}
\frac{1}{G} & = & \frac{1}{G^{(0)}} - \Sigma (k, \omega_k )
                                          \nonumber      \\
 &  \approx  &  Z^{-1}(E_c ) (\omega_k \pm  v_F k)
                                         \nonumber       \\
  &   &  + Z^{-1}(E_c )
    ( \omega_k \pm  v_F k ) \alpha (g)  \log (E_c )
\end{eqnarray}
where
\begin{equation}
\alpha (g) =  \frac{1}{4 N_s N_T }
 \left( \sqrt{1+g} + \frac{1}{\sqrt{1+g}} - 2 \right)
\label{gam}
\end{equation}

The requirement of cutoff-independence of the electron
Green function leads to absorb the logarithmic dependence on
$E_c$ into the renormalization factor $Z(E_c)$. Thus we
get the scaling equation
\begin{equation}
E_c \frac{d}{d E_c} \log Z (E_c )  =  \alpha (g).
\label{flow}
\end{equation}
We observe that $Z(E_c)$ follows a power-law behavior as a
function of the energy cutoff $E_c$. We may identify the
factor $Z(E_c)$ with the weight of the electron quasiparticles,
which vanishes as the Fermi level is approached in the limit
$E_c \rightarrow 0$. We capture in this way one of the most genuine
features of the many-body theory in 1D, namely the absence of
low-energy excitations with fermionic character, which defines
the Luttinger liquid class of 1D electron liquids.

\section{TUNNELING DENSITY OF STATES}

The power-law behavior of the quasiparticle weight $Z(E_c)$
translates into a similar dependence on energy of the tunneling
density of states $n(\varepsilon )$. By trading the dependence
on the high-energy cutoff by the energy measured with respect to
the Fermi level, we obtain the scaling behavior
\begin{equation}
 n (\varepsilon ) \sim N_s Z (\varepsilon ) \sim
    \varepsilon^{\alpha (g) }
\label{pl}
\end{equation}
Usually the behavior of the TDOS is obtained by measuring its
dependence on temperature or bias voltage. The exponent in
(\ref{pl}) is then to be compared with the measurements of
power-law behavior observed in the plots of conductance versus
temperature, or in the differential conductivity.

Regarding the experiments in Ref. \onlinecite{kprl}, we are
in disposition to explain the oscillations observed in the
exponent of the conductance, in a temperature regime
corresponding to a thermal energy of the order of the level
spacing $\Delta \varepsilon $. The point is that, as we shift
the Fermi level along the structure of single-particle peaks
in the spectrum, the $\alpha $ exponent has to reflect the
modulation of the periodic factor $N_T (\varepsilon )$. As
mentioned before, the period of the oscillations is in
agreement with the change in gate voltage required for a
shift $\Delta \varepsilon $ of the Fermi level. We comment
now on several features regarding the amplitude of the
oscillations.

\begin{figure}
\includegraphics*[width=1.0\linewidth]{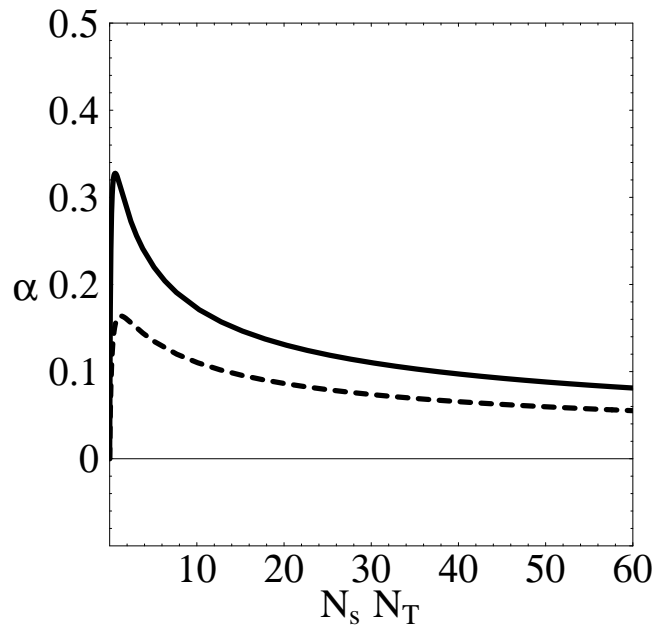}
\caption{Plot of the exponent $\alpha $ as a function
of the product of the number of subbands $N_s$ and the weight
$N_T$ in the single-particle spectrum.}
\label{alpha}
\end{figure}

It becomes very convenient to plot the expression of the
exponent $\alpha (g) $ as a function of the variable
$N_s N_T$. This depends also on the effective coupling
$\tilde{V} (q_0) e^2/v_F$ of the Coulomb interaction. We have
estimated this quantity by using a value of $q_0$ consistent
with the nanotube length, and taking into account the screening
of the interaction, mainly by other shells in the nanotube.
Thus, sensible values of $\tilde{V} (q_0) e^2/v_F$ turn out to
be of the order of $\sim 10$. The exponent $\alpha $
has been represented in Fig. \ref{alpha} as a function of
$N_s N_T$, for two different choices of the effective coupling.
We may assume the typical situation in which the MWNTs are
hole-doped, with a number of subbands crossing the Fermi level
$N_s \approx 6-10$\cite{kr}. We observe that, for such an
estimate of $N_s$, the modulation of the $N_T$ factor (with
the amplitude reported in section III) accounts for
oscillations of $\alpha $ reaching values as high as
$\approx 0.3$ (for a large value of the effective coupling)
and as low as $\approx 0.05$ (for a small value of $g$ and
a large number of subbands).

\begin{figure}
\includegraphics*[width=1.0\linewidth]{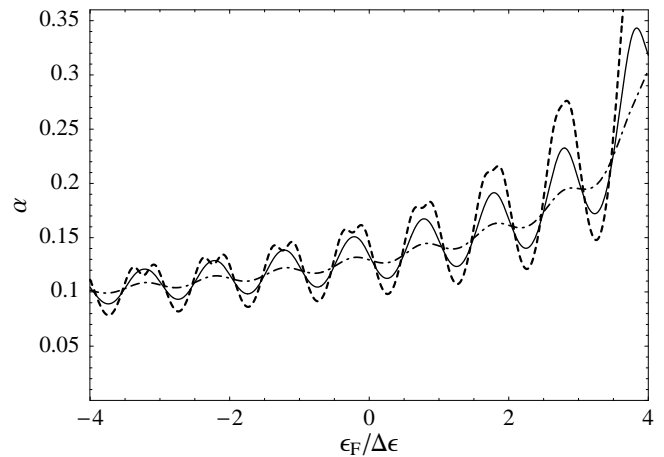}
\caption{Plot of the exponent $\alpha$ as a function of the Fermi energy $\varepsilon_{F}$ (in units
 of $\Delta \varepsilon$) for different values of temperature $kT/\Delta \varepsilon=0.1$ (dashed), 0.15 (solid), 0.3 (dotted-dashed).}
\label{osci}
\end{figure}

It is actually quite plausible that the drift observed in general
in the average value of the $\alpha $ exponent (towards larger
values for higher gate voltages) may be the consequence of a change
in the Fermi velocity as the Fermi level approaches the top of the
lowest partially filled subband. We note that the height reached
by the oscillations in the $\alpha $ exponent is different in the
two plots of experimental measures shown in Figs. 2(c) and 2(d)
in Ref. \onlinecite{kprl}. This may be attributed to a different
relative position of the Fermi level between two consecutive
subbands in the nanotube bandstructure. We have modeled the
proximity to the point with divergent density of states at the
top $\varepsilon_T$ of a given subband by allowing for a
dependence of the Fermi velocity $v_F (\varepsilon_F)  \sim
\sqrt{\varepsilon_T - \varepsilon_F}$. When this is taken into
account in conjunction with the modulation of the periodic
function $N_T(\varepsilon_F )$, we obtain a picture for the
dependence of the $\alpha $ exponent on gate voltage which is
in good qualitative agreement with the oscillations reported
in Ref. \onlinecite{kprl}, as shown in Fig. \ref{osci}.

We observe the similarity of the plot in Fig. \ref{osci} with
the experimental curves in Fig. 1 for the exponent over the 20 V
range in the gate voltage. The oscillations are significant for
low values of the thermal energy $k_B T$, as compared to
$\Delta \varepsilon$. For higher
temperatures the thermal fluctuations erase the structure of
superposed peaks in (\ref{edf}), what leads in turn to the
disappearance of the oscillations in the $\alpha $ exponent.

An important point reinforcing the above interpretation
of the experimental measures is that the depedendence on the
variable $N_s N_T(\varepsilon_F )$ shown in Fig. \ref{alpha}
translates naturally into an asymmetric behavior of the
exponent $\alpha $, when the oscillations approach
the peak of the curve near $\alpha \approx 0.3$. If we consider
for instance a doped nanotube with $N_s = 10$ and an effective
coupling $\tilde{V} (q_0) e^2/v_F \approx 12$, we observe that
the exponent may run from a value $\alpha \approx 0.18$
for $N_T = 1$ to a value $\alpha \approx 0.3$ for reasonably
small $N_T$. However, it is clear that the other side of the
oscillation does not have the same magnitude. This is consistent
with the fact that inflection points in the plots of the
conductance versus temperature have been observed for peak
values in the exponent oscillations. In our model,
the values corresponding to a dip for $N_T > 1$ do not deviate
much from the high-temperature behavior recovered for
$N_T \approx 1$. It is also remarkable that the inflection
points have been found experimentally to correspond to the
highest oscillations for large gate voltage. Again, this is
consistent with our theoretical interpretation, in which the
lower values of the exponent $\alpha $ correspond to smaller
values of the effective coupling that do not support large
variations upon changes in $N_T$, as shown from the lower curve
in Fig. \ref{alpha}.

\section{MAGNETIC FIELD EFFECTS}

The experiments reported in Ref. \onlinecite{kprl} have also
shown that the exponent in the power-law behavior of the
zero-bias conductance may depend on the applied transverse
magnetic field. The magnetic length $l = \sqrt{c /eB}$
corresponding to the field strength used in the experiments
($B = 4 \; {\rm T}$) is anyhow larger than the radius of the
outer shells in the MWNTs. Then, the resulting band structure
must be far from the development of Landau levels, although it
must already lead to the features producing the variation
observed in the $\alpha $ exponent.

To study the influence of the magnetic field, we have
analyzed the change that it may produce in the band structure of
nanotubes with different chiralities. Previous investigations have
addressed the formation of Landau subbands in carbon nanotubes
upon application of a transverse magnetic field. Here we have
undertaken a similar computational approach, but applied to
nanotubes of large radius $R \sim 10 \;{\rm nm}$, corresponding
to active shells in multi-walled nanotubes. We have thus
resorted to a tight-binding calculation of the hamiltonian in the
carbon lattice, with the usual prescription of correcting the
transfer integral by appropriate phase factors
\begin{equation}
\exp \left(i \frac{e}{c}
   \int_{\bf r}^{\bf r'} {\bf A} \cdot {\bf d l} \right)
\end{equation}
depending on the vector potential ${\bf A}$ between
nearest-neighbor sites ${\bf r}$ and ${\bf r'}$. The vector
potential has been chosen in particular with the appropriate
functional dependence at the nanotube surface, as described in
Ref. \onlinecite{sdd}.

\begin{figure}
\includegraphics*[width=1.0\linewidth]{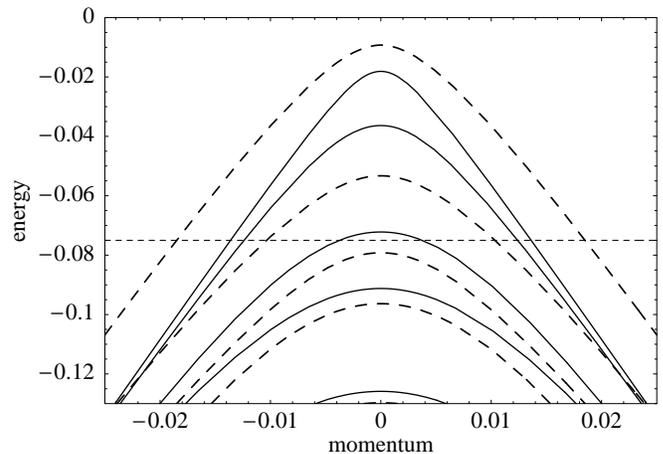}
\caption{Energy dispersion relation of a (280,0) zig-zag nanotube
with $B=0$ (solid lines) and $B=4$T (dashed lines). The position
of the Fermi level for a hole-doped nanotube is also indicated.
Energy is in eV (we used a hopping parameter $t=2.5$eV between
nearest neighbor carbon atoms) and momentum is in \AA$^{-1}$).}
\label{mag}
\end{figure}

The results of the numerical diagonalization of the
tight-binding hamiltonian for a magnetic field $B = 4 \; {\rm T}$
are shown in Fig. \ref{mag}. The band structure corresponds to
the particular case of a zig-zag nanotube, but it is quite
representative of the changes produced in chiral nanotubes upon
application of the magnetic field. There is an incipient tendency
of the subbands to flatten, an effect which is precursor of
the formation of Landau subbands at much higher magnetic field.
The two low-energy subbands spanning the gap for semiconducting
nanotubes tend to approach, as the lowest Landau level is always
at zero energy in the carbon nanotubes. The next subbands
move in the opposite direction, towards a very slow
reaccomodation along the direction of the higher Landau levels.

In this perspective, there is a significant modification in the
low-energy spectrum whenever the number of subbands crossing the
Fermi level changes, as a consequence of switching on the magnetic
field. This happens in particular when the Fermi level for $B = 0$
is slightly below the top of one of the subbands which is pushed
downwards under the action of the magnetic field. Recalling
the discussion of the preceding section, this is the instance
that corresponds to the higher gate voltages in Fig. \ref{osci},
where the value of the exponent $\alpha \approx 0.3$
is approached. In these circumstances, a couple of degenerate
subbands with large density of states are lost at the Fermi level
upon application of the magnetic field. Although the number
$N_s$ of subbands contributing to the conduction decreases, the
main effect influencing the computation of the exponent $\alpha $
comes from the larger value of $v_F$ in the remaining (outer)
degenerate subbands. Taking a value of the Fermi velocity in
correspondence with the Fermi level shown in Fig. \ref{mag},
we find that the exponent is reduced down to a value
$\alpha \approx 0.1$, in agreement with the experimental
observation reported in Ref. \onlinecite{kprl}.

\begin{figure}
\includegraphics*[width=1.0\linewidth]{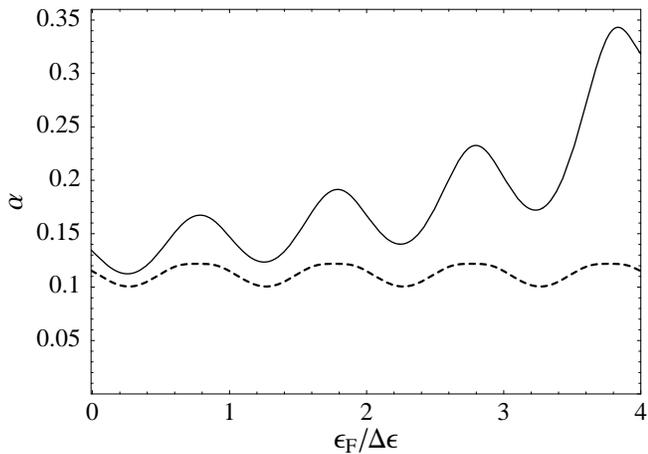}
\caption{Plot of the exponent $\alpha$ as a function of the Fermi energy $\varepsilon_{F}$ (in units
 of $\Delta \varepsilon$) for $B=0$ (solid line) and $B=4$T (dashed line). Here the temperature is
$kT/\Delta \varepsilon=0.15$.}
\label{osci2}
\end{figure}

Although no measures have been reported over a given range of gate
voltage in the presence of magnetic field, it is very instructive
to apply Eq. (\ref{gam}) to find the results from the change in
the band structure shown in Fig. \ref{mag}. The gate-voltage
dependence obtained in this way for the $\alpha $ exponent at $B =
4 \; {\rm T}$ is represented in Fig. \ref{osci2}. We observe that
the plot of the exponent is now below the curve corresponding to
the oscillations at $B = 0$, which is in agreement with the
general statement made in Ref. \onlinecite{kprl}. However, the
comparison with the experimental observation reported at a dip in
the oscillations ($\alpha \approx 0.06$ at zero magnetic field)
shows that our prediction falls short to account for the measured
reduction. The experimental result reported at $B = 4 \; {\rm T}$,
$\alpha \approx 0.005$, implies anyhow a too large effect on
transport properties, which is difficult to reconcile with the
change in the band structure for such a relatively small magnetic
field. In this respect, we point out that the evidence for a
well-defined continuous plot is not so clear for that particular
measure of the conductance as in other cases\cite{aop}. More
experimental input will be needed, over a complete range of gate
voltages, in order to confront the theoretical predictions with a
more precise picture of the magnetic field effects on the
transport properties.

\section{DISCUSSION}

Now we discuss the main distinctive features of our theoretical
approach, which lead to very good agreement with the experimental
observations.


I) The novelty of our framework reflects itself in the prediction of
the periodic dependence of the exponent $\alpha $ on the Fermi energy
of the system, due to the oscillatory behavior of the factor
$N_T(\varepsilon_F) $ as the Fermi level sweeps the structure of
single-particle levels. We estimated the value of $\Delta
\varepsilon $ in section II and inferred the variation needed in
the gate voltage $V_g$
to produce that shift in energy from the capacitive
coupling between the gate and the MWNT. We thus estimated a value of
the period for $V_g$ of the order of $\sim 2 \; {\rm V}$, in agreement
with the experimental results. As we have shown in Fig. 7, our model
can explain the oscillations in the $\alpha $ exponent, leading to plots
that are in good qualitative agreement with the experimental results
showing the dependence on the gate voltage. The precise form of these
experimental observations may change from one sample to another, but
it remains clear that the oscillations with periodicity
$\Delta V_g$ are in correspondence with the level spacing due to
the quantization of single-particle levels.

II) The oscillations in the experimental plot of $\alpha$ vs. $V_g$
were observed for values of $k_B T$ lower than a threshold which is
consistent with our estimate for the level spacing $\Delta \varepsilon$.
Furthermore, our approach explains naturally the existence of an
inflection point $T^{*}$ in the log-log plot of the conductance
versus the temperature, for values of $V_g$ corresponding to peaks in
the plot of $\alpha $. As reported in Ref. \onlinecite{kprl}, the
conductance keeps following an approximate power-law behavior below
the temperature $T^{*}$, but with an exponent significantly enhanced
with respect to the high-temperature value. In our model, this is
a natural consequence of the asymmetry in the range of variation
of $\alpha $ in the upper plot of Fig. \ref{alpha}, when the mean
value of the product $N_s N_T$ is in the range $6-10$. Peak values of
$\alpha$ as large as $\approx 0.3$ can be reached at low temperatures
(which means $N_T < 1$). However, minimum values of $\alpha$ are
reached at the other side of the oscillation, for $N_T > 1$.
In that case, low temperature does not mean a significant change
of the exponent with respect to the mean, high-temperature value.

\begin{figure}
\begin{center}
\mbox{\epsfxsize 4.2cm \epsfbox{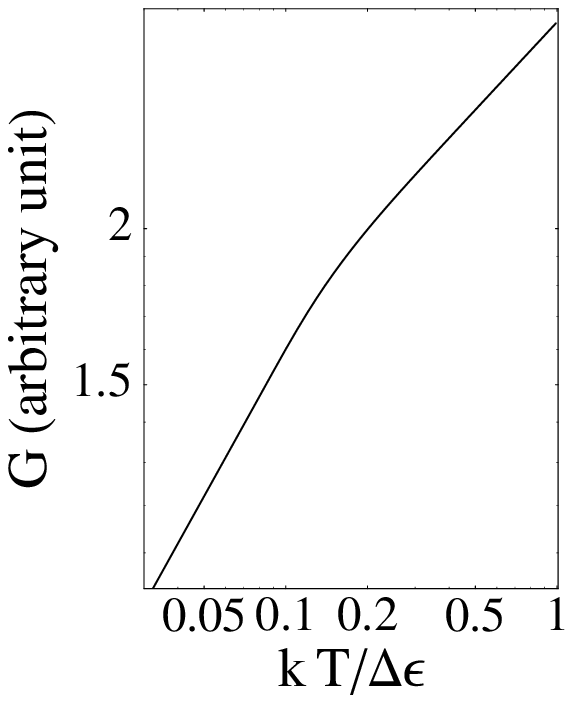}
\epsfxsize 4.2cm \epsfbox{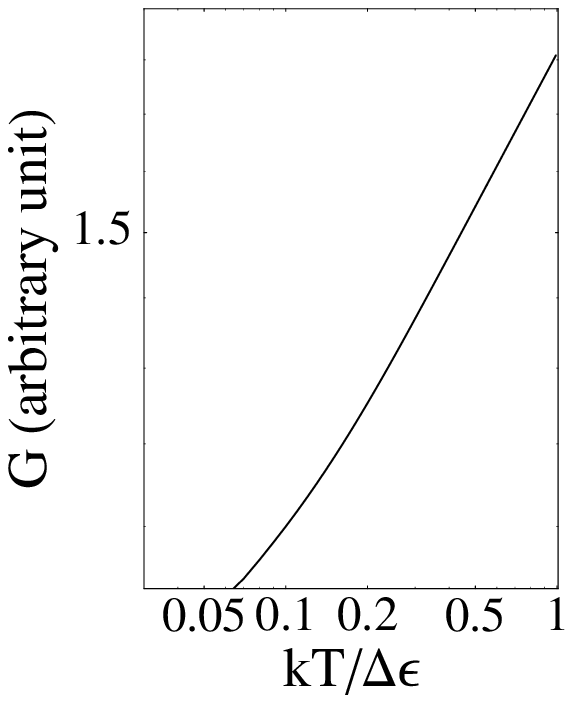}}
\end{center}
\caption{Log-log plots of the conductance as a function of the
temperature. In the left plot the Fermi energy is fixed at
$\varepsilon_{F} \approx 3.7 \Delta \varepsilon$, i.e. at the last
peak shown in Fig. \ref{osci}. In the right plot the Fermi energy
is fixed at $\varepsilon_{F} \approx - 3.7 \Delta \varepsilon$,
i.e. at the first minimum shown in Fig. \ref{osci}.}  \label{infl}
\end{figure}

Thus, it can be shown that our approach provides a sensible
description of the power-law dependence and the inflection point
in the plot of the conductance as a function of temperature. In
our framework, we may introduce the temperature as the relevant
variable by trading the scale depedence on the cutoff $E_c$ by
scale dependence on the thermal energy $k_B T$. Then we can apply
renormalization group equations like (\ref{flow}), but now taking
into account that the right-hand-side $\alpha (g)$ depends on
temperature through the factor $N_T$. We have represented in Fig.
\ref{infl} the behavior of the conductance $G$ that results from
the temperature dependence of the renormalization factor $Z(k_B
T)$. We observe that an inflection point in the log-log plots
appears only for peak values of the $\alpha $ exponent, with
values of the crossover temperature $T^*$ corresponding to $k_B
T^* \sim \Delta \varepsilon$ \cite{last}. Our description is also
consistent with the experimental fact that such inflection points
arise for gate voltages giving the largest values of $\alpha $. In
our approach, the smaller values of $\alpha $ correspond in
general to oscillations along the lower curve in Fig. \ref{alpha},
which naturally lends less sensitivity to the exponent upon
changes in the temperature variable.

\section{CONCLUDING REMARKS}

In this paper we have developed a framework that applies to an
intermediate regime between that of the Coulomb blockade and the
Luttinger liquid behavior. Our approach accounts for the main
experimental features in the zero-bias conductance reported in
Ref. \onlinecite{kprl}. This represents a significant progress
with respect to the arguments presented in that reference to
interpret the experimental observations. There it was actually
claimed that a nonconventional Coulomb blockade, arising from
tunneling into a strongly interacting disordered metal, could
be at the origin of the exponent oscillations. However,
it was also noted that such a theory does not predict power-law
behavior above the inflection temperature $T^*$. The authors
of Ref. \onlinecite{kprl} concluded that ^^ ^^ detailed
theoretical and experimental investigations would be required
for quantitative consideration".

We believe actually that, regarding transport measurements,
the role of disorder is different
than that attributed in Ref. \onlinecite{kprl} to interpret the
experimental results. The nonconventional Coulomb blockade
predicts a power-law behavior of the conductante, with
an exponent given in terms of the mean free path $l$ by
$\alpha \approx R/N_s l$. Within this theory, the oscillatory
behavior of the exponent was related to the possible modulation
of the mean free path coming from disorder. This was supported
by recalling the theoretical work of Choi {\em et al.}\cite{ch}
about the influence of different forms of disorder on the
local density of states. A closer look at this study shows,
however, that it is highly unlikely that the combined effect of
different defects may produce the desired oscillatory pattern.

From the results of Choi {\em et al.} in Ref.
\onlinecite{ch}, we note that most part of the impurities and
defects of the carbon lattice have an effect on the density of
states extending over an energy range much larger than 1 meV.
This is the energy scale consistent with the observed oscillations
in the $\alpha $ exponent. Thus, from that source of disorder we
can only expect a more or less uniform effect on the range of gate
voltage shown in the plots of Fig. 1. On the other hand, the effect
of vacancies in the lattice is to produce much narrower peaks in the
density of states, which now may give rise to features discernible
at the meV scale. We believe that this may be the main origin
of the irregularities observed in the oscillations of the $\alpha $
exponent.

Finally, we have also shown that our framework is consistent with
the experimental results reported in Ref. \onlinecite{kprl} about
the magnetic field effects in the transport properties. Using a
tight-binding approximation adapted to include the action of a
transverse magnetic field, we have
computed the change in the band structure of the thick shells in
multi-walled nanotubes. In this case, the radius of the shells
becomes comparable to the magnetic length already for a magnetic
field of $4 \; {\rm T}$. Then, we have seen that the low-energy
subbands show an incipient tendency to flatten along
Landau levels. This may induce in general further modulation
of the exponent $\alpha $, depending on the strength
of the magnetic field.

We have seen that, for high values of $\alpha $ at $B = 0$
(about $\approx 0.3$), our approach predicts a sensible reduction
of the exponent, which is in agreement with the experimental
observation. As long as the density of states at the Fermi
level is reduced upon application of the magnetic field, we
find a picture that is consistent with the statement that,
in most cases, $\alpha $ is smaller for higher magnetic
fields\cite{kprl}. However, a clarification from the experimental
point of view is required, in order to know the exceptions to
that rule. In our framework, we may envisage some instances in
which the application of the magnetic field may induce the
proximity of the Fermi level to the top of a given subband, with
the consequent increase in the density of states (and the effective
$e$-$e$ interaction). We find in this
way that sufficiently large variations in the magnetic field may
produce a modulation in the exponent $\alpha $, rather than a
monotonous trend. Our model provides anyhow definite predictions
that are susceptible of being confronted experimentally.
In this respect, more experimental input would be desirable
to discern the effects of a transverse magnetic field, that may
be quite significant for the outer shells in multi-walled
nanotubes.

\acknowledgements

\noindent J. G. acknowledges the financial support of the
Ministerio de Educaci\'on y Ciencia (Spain) through grant
BFM2003-05317. E. P. was also supported by INFN Grant 10068.


\bibliographystyle{prsty} 
\bibliography{}

\begin{thebibliography}{99}
\bibitem{foot}
For the dominant forward-scattering processes considered here,
a sensible average of the interaction is given by $q_0 \sim 1/L$.
\bibitem{aop}
One cannot discard however a nontrivial interplay between the
quantum Hall regime and the Luttinger liquid regime at large
magnetic fields. Perturbative evidence of that has been provided
for instance in Refs. \cite{BO12}.
\bibitem{last}
The form of the crossover is here slightly different to that in
the plot provided in Ref. \cite{noiprl}. This comes from the use
there of a rough approximation for the conductance, $G \sim
T^{\alpha }$, with the $T$-dependent exponent $\alpha $. The
differential renormalization given by Eq. (\ref{flow}) provides
however a more accurate description of the crossover, leading to
the plot shown to the left in Fig. \ref{infl}.
\end{thebibliography}

\end{document}